\documentclass{article}
\usepackage[T1]{fontenc} 
\usepackage[utf8]{inputenc} 
\usepackage{ismir,amsmath,cite,url}
\usepackage{graphicx}

\usepackage{color}

\def\lbd{}	

\usepackage{lineno}

\title{
CloserMusicDB: A Modern Multipurpose Dataset of High Quality Music}

\multauthor
{Aleksandra Piekarzewicz $^{1}$ \hspace{.5cm} Tomasz Sroka $^{1,2}$
\hspace{.5cm}Aleksander Tym $^1$ \hspace{.5cm} Mateusz Modrzejewski $^{1,2}$}
{
$^1$ Closer Music\\
$^2$  Institute of Computer Science, Warsaw University of Technology, Poland\\\
{\tt\small \{aleksandra.piekarzewicz, tomasz.sroka, aleksander.tym, mateusz.modrzejewski\}@closermusic.com }
}

\def\authorname{A. Piekarzewicz, T. Sroka, A. Tym and M. Modrzejewski}

\begin{document}

\maketitle
\begin{abstract}
In this paper, we introduce \textbf{CloserMusicDB}, a collection of full length studio quality tracks annotated by a team of human experts. We describe the selected qualities of our dataset, along with three example tasks possible to perform using this dataset: hook detection, contextual tagging and artist identification. We conduct baseline experiments and provide initial benchmarks for these tasks.\\

\end{abstract}
\section{Introduction}\label{sec:introduction}

Music audio datasets are essential for advancing research in fields such as music information retrieval, machine learning, and signal processing. However, the quality of available datasets varies significantly, often due to copyright restrictions and limited access to high-fidelity recordings \cite{gui2024adapting}. This variability can affect the reproducibility and potential for generalization of research findings, underscoring the need for more comprehensive and high-quality music datasets.

We introduce CloserMusicDB, a dataset designed to provide high-quality, copyright-compliant music for research purposes along with selected subsets of metadata provided by human experts. Alongside the audio, the dataset provides \emph{hook} start and end annotations, a subset of relevant \emph{hashtags} and artist identifiers for each audio file. By ensuring consistent quality and accessibility, CloserMusicDB aims to support reproducible research and foster advancements in music-related machine learning tasks.

\section{Example tasks}
\subsection{Hook detection}
A \emph{hook} in popular music refers to a short, memorable musical or lyrical phrase designed to catch the listener's attention. It is often repeated throughout the song and can appear in various sections, such as the chorus, intro, or bridge. The hook is crafted to be immediately recognizable, driving listener engagement and retention. Its purpose is to enhance recall and it often contributes significantly to a song's commercial success. 

Hook detection refers to identifying the most prominent and repeated segments of a song, often those intended to capture the listener's attention. Hooks may vary in instrumentation and energy level, making them challenging to detect. Our dataset contains hook annotations manually labeled by expert musicians and producers, who identified these sections in each track, ensuring accurate identification of these sections.

\subsection{Contextual tagging}
In music tagging, the concept of tags involves assigning descriptive labels to tracks, albums, or artists to categorize and organize music based on its characteristics. These tags can include genre, mood, tempo, instrumentation, or specific qualities such as "energetic" or "melancholic." Tags serve to facilitate search, recommendation, and discovery of music by highlighting its intrinsic qualities \cite{won2021music}.

However, while many tags focus on musical or emotional attributes, relatively few describe the contextual or functional aspects of music—such as occasions, purposes, or specific scenarios in which the music may be used. These \emph{"hashtag"} qualities might include tags like "workout," "travel," or "background for video," which go beyond musical properties to signal practical use cases or associations. Expanding the tagging system to include such contextual tags could enhance the utility of music recommendation systems, as it would allow users to find tracks not only based on how they sound, but also how they fit specific activities or multimedia content.

\subsection{Artist identification}
Artist identification focuses on recognizing distinctive audio patterns that are characteristic of a specific artist. Our dataset contains multiple examples from the same artist, enabling models to detect consistent features and stylistic elements unique to each artist. This allows for the identification of recurring patterns across works, facilitating the differentiation between artists based on their individual musical traits.

\begin{figure*}[hbt!]
    \centering
    \includegraphics[width=\textwidth]{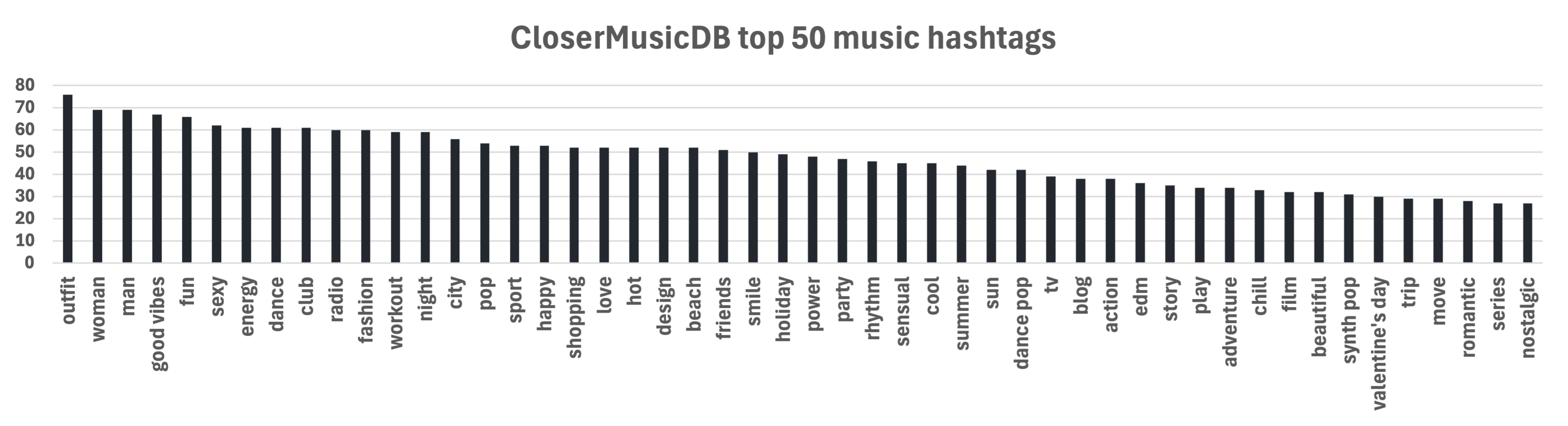}
    \caption{Fifty most frequent \textit{music hashtags} in the CloserMusicDB.}
    \label{fig:top50tags}
\end{figure*}

\section{Dataset structure}
The CloserMusicDB dataset features 106 high quality, full-length tracks (around 5 hours of audio). All tracks are recorded, produced, mixed and mastered by professionals. The files are stored as uncompressed stereo WAVs, with sampling rate of 44100 Hz and 16-bit depth. The metadata contains the following fields:
\begin{itemize}
    \itemsep0em
    \item a name of the WAV file,
    \item main artist and featured artist identifiers,
    \item timestamps of hook start and end,
    \item beats per minute (BPM),
    \item a subset of music hashtags.
\end{itemize}

There are 280 unique \textit{music hashtags} in the dataset. Fifty most frequent are presented in Figure \ref{fig:top50tags}. Most of the provided tags are contextual and indicate the potential fit of the tracks for other multimedia content, activities or occasions. Furthermore, some of the tags also describe the genre, emotions or moods associated with the track.

\section{Initial benchmarks}
\subsection{Hook detection}

We find MSAF \cite{Nieto2016SystematicEO} is able to recognize segments we annotated as hooks as individual, stand-alone sections of the tune while using Ordinal LDA boundary detection algorithm. The accuracy is 41.5\% with $\pm5s$ start/end time tolerance and 35.8\% with $\pm3s$ tolerance.

\subsection{Contextual tagging and artist identification}
We conduct baseline experiments for contextual tagging and artist identification with a transfer learning approach similar to \cite{choi2017transfer}. We extract \texttt{OpenL3} \cite{cramer2019look} embeddings for the classification tasks. We use the \emph{256mel} model type, 512 embedding length and 1s hop size. 

For both classification tasks, we use the same general model architecture -- a multi-layer perceptron with three hidden layers, 256 neurons each. We use the ReLU activation function in the hidden layers. For contextual tagging the last activation function is sigmoid, as it is a multilabel classification task. For artist identification we use a softmax, as it is a multiclass classification task. 
We use 5-fold cross validation for both tasks. For artist identification the folds are stratified based on artists and grouped by songs in order to prevent leaking embeddings from one song between splits. We train each fold for 50 epochs with the Adam optimizer \cite{kingma2017adammethodstochasticoptimization}, a learning rate of $1e-4$ and a batch size of 32. We decide on these parameters upon some initial experimentation, but perform no large hyperparameter sweep nor any data augmentations. We perform all experiments on a single RTX 4090 GPU.

In contextual tagging task we achieve $0.2998\pm0.0767$ Jaccard score (weighted by each label occurences) and $0.6772\%\pm0.0625$ ROC AUC averaged across all folds. For comparison, random guessing would yield a $0.0$ Jaccard score and $0.5$ ROC AUC.

In artist identification task we achieve $60.22\%\pm5.14$ accuracy and $0.5141\pm0.0849$ $F_1$ score averaged across all folds. In contrast, picking the most common artist would yield approximately $32\%$ accuracy and $0.1$ $F_1$ score.

\section{License}
The dataset is available under the Non-Commercial Research Community License (NC-RCL). Researchers are permitted to use the provided audio material as long as it is done only for non-commercial purposes within the research community. The redistribution of the metadata and audio tracks is conducted solely through the Zenodo platform, with a redirecting repository available on GitHub.\footnote{https://github.com/closermusic/CloserMusicDB} To request access to these files, you'll need to fill out a short form.

\section{Conclusions}
We introduce CloserMusicDB, a high-quality, human-annotated dataset of full-length studio tracks, designed to address the limitations of existing music datasets. It facilitates research in tasks such as hook detection, contextual tagging, and artist identification, with initial benchmarks demonstrating its potential for advancing music information retrieval. By providing consistent metadata and expert annotations, CloserMusicDB supports reproducible research and aims to bridge the gap between academic studies and real-world music applications.

\bibliography{ISMIRtemplate}

\end{document}